\documentclass{aa}
\usepackage{psfig}
\begin{document}
\title{Mid-Infrared Imaging of a Young Bipolar Nebula in the S287 Molecular Cloud
\thanks{Based on observations collected at the European Southern Observatory, 
        Chile (program ID 60.A-9269(A)).}}

\author{M.E. van den Ancker}
\institute{
European Southern Observatory, Karl-Schwarzschild-Str. 2, D--85748 
 Garching bei M\"unchen, Germany\\ e-mail: {\tt mvanden@eso.org}}

\date{Received <date>; accepted <date>} 

\abstract{We present diffraction-limited images in the 11.3~$\mu$m PAH 
band and in the continuum at 11.9~$\mu$m of the S287B star forming region 
obtained with the newly commissioned mid-infrared imager/spectrometer 
VISIR at the VLT.  In both filters five point-like sources are detected, 
of which at least one has a spectral energy distribution reminiscent 
of a Lada Class I source.  We also report on the discovery of a new 
Herbig Ae star in this region.  It is particularly striking that the 
brightest mid-infrared source in this region has not been detected at shorter 
wavelengths; it is a good candidate to be the driving source of the 
bipolar molecular outflow and optical reflection nebulosity previously 
reported in S287B.
\keywords{Circumstellar matter -- Stars: formation -- 
          ISM: S287B -- Reflection nebulae -- Infrared: Stars}
}
\maketitle

\section{Introduction}
\begin{figure*}
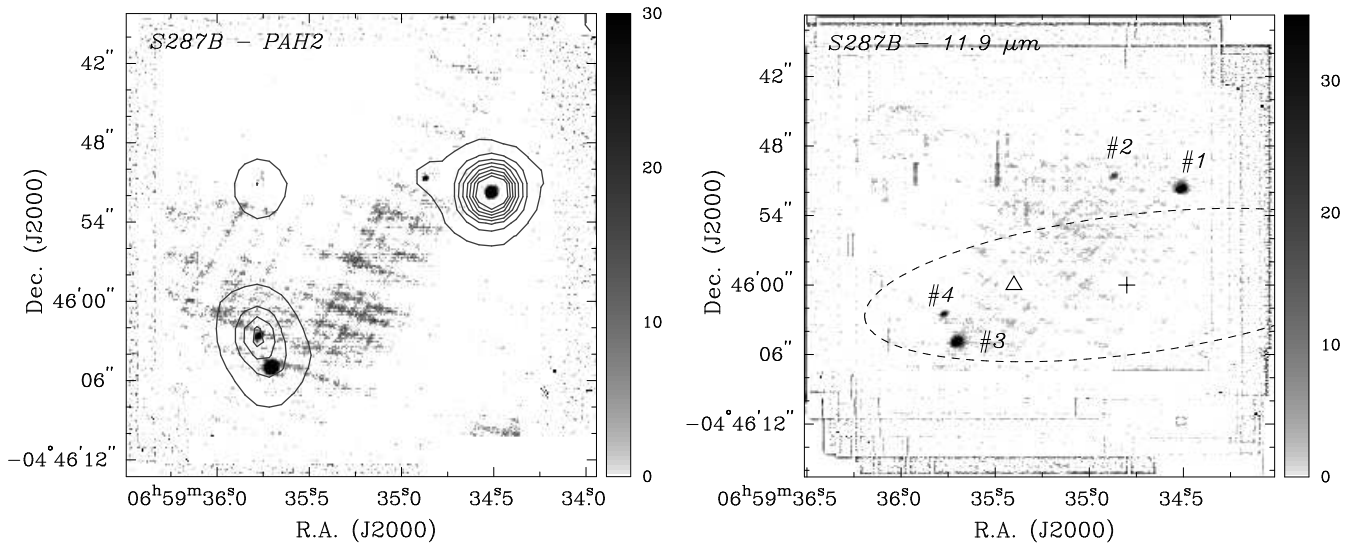

\centerline{\psfig{figure=S287B_PAH2.ps,width=8.7cm,angle=270}
            \hspace*{0.1cm}
            \psfig{figure=S287B_PAH2_2.ps,width=8.7cm,angle=270}}
\caption[]{Greyscale images of the S287B region in the PAH2 (11.3~$\mu$m; 
           left) and 11.9~$\mu$m continuum (right) filters. The overlay in 
           the PAH2 image show contours of the $K_s$ (2.2~$\mu$m) image from
           the 2MASS survey.  In the 11.9~$\mu$m continuum image we also 
           show the IRAS position (cross) and error ellipse (dashed curve) 
           of IRAS 06571$-$0441.  The position of MSX G218.1025$-$00.3638 
           is marked with a triangle.  Sources \#1--\#4 are also labelled in 
           this image.  Source \#5 is located south of the area shown here; it 
           is only detected in the chopped (negative) beam.  The low-level diagonal 
           ``striping'' pattern seen in the background of both images is an artifact 
           caused by imperfections in the subtraction of thermal emission from the 
           sky and the telescope support structure.}
\end{figure*}
The formation of massive stars is currently one of the most 
hotly debated topics in astrophysics.  Massive stars reach 
the Zero-Age Main Sequence (ZAMS) while still deeply embedded 
in their infalling envelope.  One would expect the strong 
radiation pressure from the newly formed star to clear its 
natal envelope when the star has reached a mass of 
$\sim$ 10 M$_\odot$, halting the further build-up of mass 
(e.g. Stahler et al. 2000).  
Yet we observe stars that are apparently more massive than 
this limit.  Two mechanisms 
have been proposed to circumvent this dilemma and allow the 
creation of stars more massive than 10 solar masses: mass 
build-up through disk accretion, and mergers of two or more 
less massive stars in dense stellar clusters.  It is 
currently not clear which of these two mechanism is 
responsible for forming massive stars within our galaxy 
(e.g. Bonnell \& Bate 2002; Yorke \& Sonnhalter 2002; 
Bally \& Zinnecker 2005).

To tackle this problem observationally, one would like to 
study the earliest stages of the formation of a massive star.  
Since these stars are still heavily obscured by their 
infalling envelope ($A_V$ ranging from 20 to $>$ 1000), the 
mid-infrared is the most natural wavelength regime for this 
type of study.  Of particular interest may be the small group 
of Young Bipolar Nebulae, in which the strong outflow from 
the embedded source has created a bipolar cavity, allowing 
some of the UV-visual radiation from the central star to 
escape the system (see e.g. Staude \& Els\"asser (1994) for 
a comprehensive review).  Young Bipolar Nebulae have been 
relatively well studied at optical to near-infrared wavelengths.  
However, these wavelengths are dominated by scattered light 
from the nebula, offering little information about the processes 
occurring in the innermost parts.  In contrast, the mid-infrared 
continuum is able to probe regions close to the central object, 
and reveal the true location of the illuminating source.  Moreover 
mid-infrared emission is able to penetrate large sections of the 
bipolar nebula, tracing the inner surface of the biconical cavity 
created by the central source.

In this {\it letter} we aim to contribute to the problem of 
the formation of massive stars by 
focussing on a little-studied bipolar nebula located in the 
molecular cloud associated with the H\,{\sc ii} region S287. 
The S287 molecular cloud ($l$ = 218.10$^\circ$, $b$ = $-$0.37$^\circ$, 
$d$ = 2.3 kpc; Williams \& Maddalena 1996) is an elongated structure, 
extending over about 1.5$^\circ$ 
parallel to the galactic equator. It consists of two main components 
of similar sizes and CO column densities, which are connected by a 
tenuous bridge.  The northern component harbours its densest core, 
associated with the bipolar nebula NS14.  The faint H\,{\sc ii} 
region S287 is associated with the other component.   S287 is a 
likely location of on-going star formation.  
The S287 cloud and another nearby cloud mapped by Maddalena \& Thadeus 
(1985) may all have been part of a larger molecular cloud complex 
that included G216$-$2.5 that has been partially dissociated by 
the S287 OB stars.

The intermediate-luminosity IRAS source IRAS 06571$-$0441 is 
located in the vicinity of a small reflection nebulosity at the 
northern border of S287. The region is 
associated with a bipolar outflow mapped in CO and traced by a 
string of Herbig-Haro objects (Neckel \& Staude 1992). 
The IRAS source is located in the vicinity of a likely 
young bipolar nebula.  Deep optical images by Neckel \& Staude (1992) 
reveal two regions of reflection nebulosity separated by a dark 
lane ($A_V$ $>$ 20 mag).  Its bolometric luminosity of $\approx$ 440 L$_\odot$ 
suggests the embedded source to be a massive (possibly Herbig Be) YSO.  
Although no definite optical counterpart was detected, Neckel \& Staude (1992) 
found that a substantial fraction of the optical light scattered by the 
reflection nebulosities consists of broad H$\alpha$ emission 
with pronounced P-Cygni profiles, in line with the emission 
originating in a Herbig Ae/Be or T Tauri wind.

In this {\it letter} we present high-spatial resolution mid-infrared 
images of the vicinity of the bipolar nebula associated with 
IRAS 06571$-$0441.  We report the detection of a likely counterpart 
for the embedded source, as well as several less luminous mid-infrared 
sources.  We conclude that the S287B region is a likely site of still 
on-going massive star formation.

\section{Observations}
\begin{table*}
\caption[]{Detected sources in S287B.}
\tabcolsep0.11cm
\begin{flushleft}
\begin{tabular}{cccccl}
\hline\noalign{\smallskip}
No. & R.A. (J2000)  & Dec. (J2000)  & $S_{PAH2}$ [mJy] & $S_{11.9~\mu m}$ [mJy] & 2MASS ID\\
\noalign{\smallskip}
\hline\noalign{\smallskip}
\# 1 & 06:59:34.52 & $-$04:45:51.8 & 418 $\pm$ 12 & 376 $\pm$ 21 & 06593451$-$0445517\\
\# 2 & 06:59:34.87 & $-$04:45:50.7 &  19 $\pm$  4 &  37 $\pm$  4 & 06593481$-$0445506\\
\# 3 & 06:59:35.71 & $-$04:46:05.0 & 465 $\pm$  9 & 443 $\pm$  9 & \\
\# 4 & 06:59:35.77 & $-$04:46:02.6 &  52 $\pm$  2 &  68 $\pm$  5 & 06593576$-$0446025\\
\# 5 & 06:59:35.99 & $-$04:46:15.2 &  22 $\pm$ 12 &  37 $\pm$  5 & \\
\noalign{\smallskip}
\hline
\end{tabular}
\end{flushleft}
\end{table*}
Mid-infrared images of S287B in the PAH2 ($\lambda_c$ = 11.26~$\mu$m, 
$\Delta \lambda$ = 0.59~$\mu$m) filter and in the continuum filter 
centered at 11.88~$\mu$m (PAH2\_2, $\Delta \lambda$ = 0.37~$\mu$m) 
were taken on November 27, 2004 as part of Science Verification 
of the newly installed {\it VLT Imager and Spectrometer for mid 
Infrared}\footnote{http://www.eso.org/instruments/visir/} (VISIR) 
at the {\it Very Large Telescope} (VLT).  VISIR uses a DRS 
(former Boeing) 256 $\times$ 256 BIB detector.  The pixel scale was 
set to 0\farcs127, resulting in a 32.5\arcsec\ $\times$ 32.5\arcsec\ 
field of view.  Subtraction of the 
thermal emission from the sky, as well as the telescope itself, was 
achieved by chopping in the North-South direction with a chop throw of 
20\arcsec, and nodding the telescope in the opposite direction with 
equal amplitude.  The cosmetic quality of the images was further 
improved by superimposing a random jitter pattern (with maximum throw 
2\arcsec) on the nodding sequence, so as to minimize the effect of bad 
pixels in the detector array on the final science data. 
Total integration times were 15 and 30 minutes 
for the PAH2 and the 11.9~$\mu$m continuum filter, respectively.
  
The mid-infrared standard stars HD~22663 (K1 III) and 
HD~59311 (K5 III), observed just before and 
after the science observation, were used to flux calibrate the 
data and as point spread function (PSF) reference stars.  
The flux calibration obtained by using one or the other of these 
two standard stars differs by about 2\%.  For our final 
calibration we have adopted an average of both calibrations, 
and we adopt a 2\% systematic error on the absolute flux 
calibration based on this difference. 
Although the seeing at the time of observation was 0\farcs9 in 
the optical, the FWHM of our standard star images is only  
0.35\arcsec, comparable to the diffraction limit of the 
VLT at 11.3~$\mu$m (0.28\arcsec).

After a standard data reduction consisting of co-adding of frames, 
flat fielding and bad pixel removal,  the chopped and nodded 
images of S287B were combined into one single frame 
per filter, shown in Fig.~1.  In both filters, we detect five 
point-like sources.  In the remainder of this paper we will refer 
to these sources as \#1 to \#5, in order of increasing right ascension.  
Note that source \#5 was only detected in the chopped (negative) beam; 
it is located South of the area shown in Fig.~1.  All five detected 
sources appear point-like. By comparison with our PSF 
reference stars, we derive upper limits of $<$ 0\farcs15 on 
the spatial extent of these compact sources.  No evidence for 
extended emission could be detected in any of our mid-infrared 
images.  Synthetic aperture 
photometry was performed on all five point sources using a circular 
aperture of 15 pixels (2\farcs0) diameter, with a 5 pixel wide 
annulus surrounding this aperture used to subtract any background 
residuals.  Different choices of the radii did not significantly 
alter our results.  The resulting new mid-infrared photometry of 
the five detected sources in S287B is listed in Table~1.  Listed 
errors include contributions due to photon noise, residuals from 
the background subtraction, and pixel gain variations (which 
are strongly reduced due to our choice of an observing strategy 
including dithering), as well as the 2\% error in the absolute 
flux calibration.
%

Although the relative astrometry of the VISIR data is highly 
accurate, the absolute positions derived from the data is 
subject to an uncertainty -- mainly due to the finite 
accuracy of the telescope pointing -- of a few arcseconds. 
A careful inspection of our images shows that sources \#1, \#2 
and \#4 can be associated with a counterpart that has been 
detected in the $K_s$ (2.17~$\mu$m) band by the 2MASS survey 
(Fig.~1).  We have therefore registered our images using 
the 2MASS coordinates of source \#1 (2MASS 06593451$-$0445517). 
Subsequently, the positions of all five detected point sources 
were determined by the fitting of a Gaussian to each source.  
In Table~1 we also report the so derived position for each source.  
We estimate this astrometry to be accurate at the 0\farcs1 level.

\section{Analysis and discussion}
\begin{figure}
\centerline{\psfig{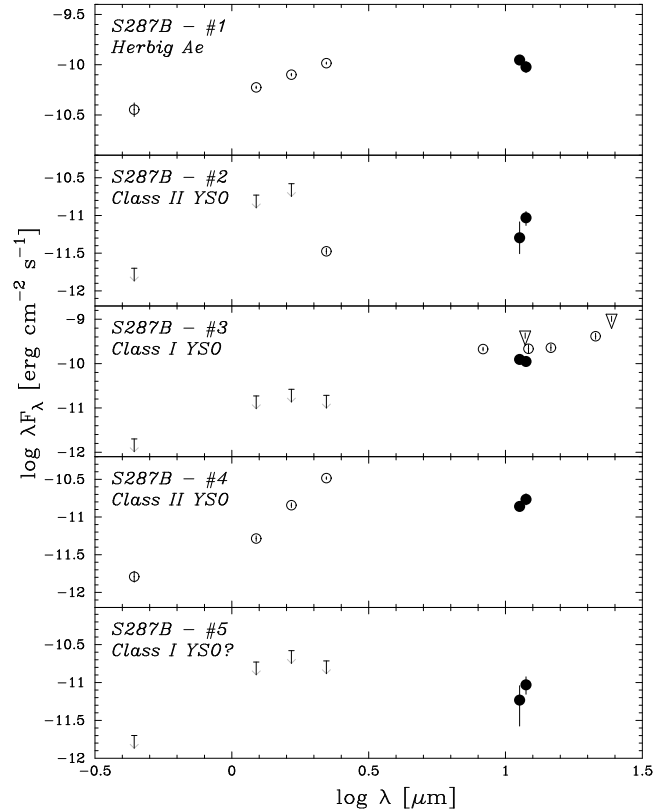}}
\caption[]{Spectral Energy Distributions (SEDs) of all five 
           detected sources in the S287B region. Open circles 
           show data from the GSC2.2, 2MASS and MSX catalogues.  
           The new VISIR photometry is shown as filled circles.
           IRAS data are indicated by triangles. 
}
\end{figure}
An inspection of images from the 
{\it Digitized Sky Survey}\footnote{http://www-gss.stsci.edu/Dss/dss\_home.htm} 
(DSS) reveals likely optical counterparts for our sources \#1 and \#4.  
Based on these identifications, we have constructed optical-infrared 
Spectral Energy Distributions (SEDs) of the five sources detected in 
our VISIR data (Fig.~2).  Upper limits to the near-IR and optical fluxes 
were derived from the 2MASS and DSS data and are plotted here as well.

Our source \#1 -- identical to star 1 in the field of IRAS 06571$-$0441 
in the nomenclature of Neckel \& Staude (1992) -- is surrounded by a 
small reflection nebula in the DSS images.  No trace of this 
nebulosity was recovered at the mid-infrared wavelengths samples by 
our VISIR data. Source \#2 may be identified with the faint feature b 
in the field of IRAS 06571$-$0441 detected by Neckel \& Staude (1992).  
Sources \#3 and \#4 are located 
close to the nebulous knot of material (f in the terminology of 
Neckel \& Staude) they identified as a Herbig-Haro object.

The relative faintness of source \#2 in our mid-infrared images 
clearly shows that it is not the optical counterpart of IRAS 06571$-$0441, 
as suggested by Neckel \& Staude.  The {\it IRAS Point Source Catalogue} (v. 2) 
positional error ellipse of IRAS 06571$-$0441 is instead compatible with our 
source \#3 or source \#4 (Fig. 1).  Based on its brightness in the 
VISIR images, and the lack of an associated 2MASS source -- 
indicative of a very red energy distribution -- we identify 
our source \#3 as the likely counterpart of IRAS 06571$-$0441.  
We note, however, that its 11.9~$\mu$m flux in our VISIR images 
is significantly lower than the IRAS 12~$\mu$m flux; likely also 
extended mid-infrared below our detection limit is present.  
Curiously, the position of the counterpart of IRAS 06571$-$0441 
detected by the {\it Midcourse Space Experiment} (MSX; Egan et al. 2003) 
does not correspond to any of the sources in our VISIR image; the 
presence of extended mid-infrared emission not detected in our images 
could possibly also explain this discrepancy.   

Optical spectroscopy by Neckel \& Staude (1992) has shown that star \#1 is 
an early A-type star showing strong H$\alpha$ emission.  After correcting 
our photometry for foreground extinction adopting $A_V$ = 3.4 mag (Neckel \& 
Staude 1992), and the normal (i.e. $R_V$ = $A_V/E(B-V)$ = 3.1) interstellar 
extinction law by Cardelli et al. (1989), we derive the bolometric luminosity 
of star \#1 to be around $L$ = 55~L$_\odot$, or about 20 times 
less luminous than the IRAS source.  Our new mid-IR photometry reveals 
a strong mid-IR excess above levels that can be explained by the stellar 
photosphere in source \#1  (Fig.~2).  We conclude that 
this object is a likely new Herbig Ae star.  The possible presence 
of intrinsic PAH emission at 11.3~$\mu$m from this object 
(Table 1) would suggest that it is a Meeus et al. (2001) group I 
source, i.e. it possesses a flared disk (Acke \& van den Ancker 2004). 
Assuming an effective 
temperature of 8900~K (corresponding to a spectral type of A2; Schmidt-Kaler 
1982), and the above-mentioned luminosity of 55~L$_\odot$, we derive a 
mass of 2.5--3.0~M$_\odot$ and an age of 2--5 Myr for this object 
by comparing its position in the Hertzsprung-Russell diagram with the 
pre-main sequence evolutionary tracks by Palla \& Stahler (1993).

Although less information is available on source \#2 and source \#4, 
they also have energy distributions reminiscent of many class II YSOs.  
Their bolometric luminosities of 13 and 17 L$_\odot$, respectively, 
suggest that they may also be intermediate-mass young stars.

Source \#3 and source \#5 do not have a counterpart at 
wavelengths $<$ 10~$\mu$m.  Source \#3 (the likely counterpart 
of IRAS 06571$-$0441) is by far the most luminous object in the region 
($L_{\rm bol}$ $\approx$ 1.3 $\times$ 10$^3$ L$_\odot$).  It is also 
very red: it has the energy distribution of a Lada Class I source.  
We conclude that source \#3 appears 
to be a particularly good candidate for being the driving source for 
the bipolar outflow previously detected in molecular material.  If 
the circumstellar material is distributed appropriately around the 
central star (e.g. in a disk seen close to edge-on), the star -- 
although not visible in the optical from our vantage point -- could also 
well be the dominant source illuminating the nebulosities in the field. 
The geometry of star \#3 would be typical of that of a ``bipolar nebula'' 
with disk seen edge-on (c.f. Staude \& Els\"asser 1993).

\section{Conclusions}
Our new VISIR observations have revealed five young stars in the 
part of S287 that was sampled by our data.  In particular our  
mid-infrared data have revealed the driving source of the molecular 
outflow and the bipolar nebula seen in scattered light to be an 
embedded massive young star whose energy distribution is 
reminiscent of a Lada Class I source.  We have also shown S287B 
to be associated with a likely new Herbig Ae star.  The presence of 
presumably very young Class I objects in close vicinity to somewhat 
more evolved Class II objects is noteworthy.  It certainly raises 
some questions about the scenario for star formation in this region: 
either not all sources are contemporaneous, or the class I (active 
accretion) phase must have lasted significantly longer in some objects 
in this regions than in others.  Further research is necessary to 
clarify this issue.

It is also striking to note that the most luminous source in this 
region had not been detected at shorter wavelengths, nor had it been 
recognized previously from lower spatial resolution infrared data. 
This offers a candid illustration of the new insights that can be 
gained by the increase in spatial resolution in the mid-infrared that 
the combination of mid-infrared cameras on large-aperture ground-based 
telescope brings.

\acknowledgements{
The author would like to thank the anonymous referee, whose comments 
improved both contents and presentation of the manuscript. 
This publication makes use of data products from the Two Micron All Sky 
Survey, which is a joint project of the University of Massachusetts and 
the Infrared Processing and Analysis Center, funded by the National 
Aeronautics and Space Administration and the National Science Foundation.}


\begin{thebibliography}{}
\bibitem{}
Acke B., van den Ancker M.E., 2004, A\&A 426, 151
\bibitem{}
Bally J., Zinnecker H., 2005, AJ 129, 2281
\bibitem{}
Bonnell I.A., Bate M.R., 2002, MNRAS 336, 659
\bibitem{}
Cardelli J.A., Clayton G.C., Mathis J.S., 1989, ApJ 345, 245
\bibitem{}
Egan M.P., et al., 2003, ``The Midcourse Space Experiment Point 
 Source Catalog version 2.3'', Air Force Research Laboratory Technical 
 Report AFRL-VSTR 2003-1589.
\bibitem{}
Maddalena R.J., Thaddeus P., 1985, ApJ 294, 231
\bibitem{}
Meeus G., Waters L.B.F.M., Bouwman J., van den Ancker M.E., Waelkens C., 
 Malfait K., 2001, A\&A 365, 476 
\bibitem{}
Neckel T., Staude H.J., 1992, A\&A 254, 339
\bibitem{}
Palla F., Stahler S.W., 1993, ApJ 418, 414
\bibitem{}
Schmidt-Kaler Th., 1982, Land\"olt-Bornstein Catalogue VI/2b
\bibitem{}
Staude H.J, Els\"asser H., 1994, Astron. Astrophys. Review 5, 165
\bibitem{}
Stahler S.W., Palla F., Ho P.T.H., 2000, in {\it ``Protostars and 
 Planets IV''}, Eds. V. Mannings, A.P. Boss \& S.S. Russell, Univ. of 
 Arizona Press, Tucson, p. 327
\bibitem{}
Williams J.P., Maddalena R.J., 1996, ApJ 464, 247
\bibitem{}
Yorke H., Sonnhalter C., 2002, ApJ 551, 461
\end{thebibliography}
\end{document}